\documentclass[twocolumn,prb,amsmath,amssymb,floatfix,superscriptaddress,showpacs]{revtex4-1}

\usepackage{times}
\usepackage{color}
\usepackage{graphicx}

\usepackage{dcolumn}
\usepackage{bm}

\begin{document}

\title{Electric field effect on short-range polar orders in a relaxor ferroelectric system}

\author{Zhijun~Xu}

\affiliation{NIST Center for Neutron Research, National Institute of Standards and Technology, Gaithersburg, Maryland 20877, USA}
\affiliation{Department of
    Materials Science and Engineering, University of Maryland, College Park,
    Maryland, 20742, USA}

\author{Fei Li}
\affiliation{Department of Materials Science and Engineering, Materials Research Institute, Pennsylvania State University, University Park, Pennsylvania 16802, USA}
\affiliation{Electronic Materials Research Laboratory, Key Laboratory of the Ministry of Education and International Center for Dielectric Research, Xi’an Jiaotong University, Xi’an 710049, China}

\author{Shujun Zhang}
\affiliation{Institute for Superconducting and Electronic Materials, Australian Institute of Innovative Materials, University of Wollongong, Wollongong, New South Wales 2500, Australia}

\author{Christopher Stock}
\affiliation{School of Physics and Astronomy, University of Edinburgh, Edinburgh EH9 3JZ, United Kingdom}

\author{Jun Luo}
\affiliation{TRS Technologies Inc., 2820 E College Avenue, State College, PA 16801}

\author{Peter M. Gehring}
\affiliation{NIST Center for Neutron Research, National Institute of Standards and Technology, Gaithersburg, Maryland 20877, USA}

\author{Guangyong~Xu}
\affiliation{NIST Center for Neutron Research, National Institute of Standards and Technology, Gaithersburg, Maryland 20877, USA}



\begin{abstract}

Short-range polar orders in the relaxor ferroelectric material PbMg$_{1/3}$Nb$_{2/3}$O$_3$-$28\%$PbTiO$_3$ (PMN-28PT) have been studied using neutron diffuse scattering. An external electric field along [110] direction
can affect the diffuse scattering in the low temperature rhombohedral/monoclinic phase.
Diffuse scattering intensities
associated with [110] short-range polarizations are partially suppressed, while those arising from [1$\bar{1}$0]
polarizations are enhanced. On the other hand, short-range polar orders along other equivalent $\langle110\rangle$ directions, {\it i.e.}  [101],[10$\bar{1}$], [011], and [01$\bar{1}$] directions, are virtually unaffected by the field. Our results, combined with previous work,
strongly suggest that most part of short-range polar orders in PMN-$x$PT relaxor systems are robust
in the low temperature phase, where they couple strongly to ferroelectric polarizations of the surrounding ferroelectric domains, and would only respond to an external field indirectly through ferroelectric domain rotation.

\end{abstract}

\maketitle

PbMg$_{1/3}$Nb$_{2/3}$O$_3$ (PMN) is a prototypical lead based relaxor system where no long-range polar order
can be established without an external electric field~\cite{PZT1,Uchino,Service,Shujun2012}. Due to the existence of strong random field~\cite{Random_Field_Ori,Random_Field_FE,Random_Field0,Random_Field1,Random_Field2}, short-range polar orders (SRPO), or, sometimes referred to as "polar nano-regions" (PNR) start to appear in the system at the Burn's temperture T$_d \sim 620$~K and grow with cooling~\cite{Burns}. These SRPO are believed to contribute to many unique properties of the relaxor material~\cite{Cross}. When mixed with the classical ferroelectric PbTiO$_3$ (PT), spontaneous ferroelectric polarizations can start to develop, and the solid solutions of PMN-$x$PT naturally exhibit combined ferroelectric and relaxor  characteristics.  In the mean time, the system shows extraordinary piezoelectric properties when approaching the Morphotropic Phase Boundary (MPB). There is evidence that the SRPO are important for the high piezoelectric responses as well~\cite{PMN_Critical,FeiLi2016,FeiLi2018}. The relaxor properties gradually disappear when the system crosses MPB into the regime of classic ferroelecrtrics~\cite{Kiat,Universal_phase,PMN_phase,PMN-60PT}.  For those with low PT concentrations, i.e. on the left side of the MPB, the SRPO can actually persist into the low temperature long-range  ferroelectric ordered phase~\cite{Xu_new,Xu_coexist},  and therefore offers a fascinating platform to study how long- and short-range polar orders coexist and compete~\cite{Xu_JPSJ}.

\begin{figure}
    \includegraphics[width=0.45\linewidth,angle=90]{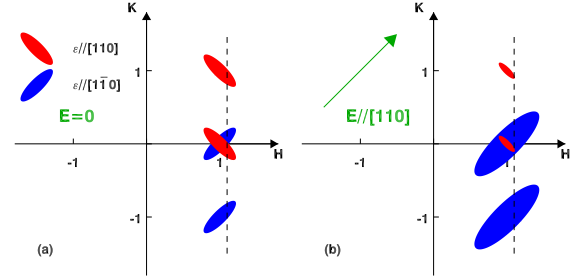}
    \caption{Schematic of the diffuse scattering intensity distribution in the (HK0) plane under (a) zero field, and (b) a E-field along [110] direction. The red and blue "wings" are intensities associated with mostly [110] and [1$\bar{1}$0] polarizations, respectively. The dashed lines indicate the locations of the (1.1,0,K) linear scans described in the text. }\label{fig:1}
\end{figure}

Diffuse scattering measurements are very sensitive to inhomogeneities in various materials systems, and have been used extensively to probe the SRPO in relaxors~\cite{PMN_xraydiffuse,PMN_diffuse,PMN_diffuse2,PMN_diffuse3,PZN_diffuse,PZN_diffuse2,PZN_diffuse3,HIro_diffuse,Xu_diffuse,PMN_efield,Matsuura,Xu_3D}. In general we find that in lead-based relaxors such as PMN-$x$PT and their analogue PZN-$x$PT, the diffuse scattering is dominated by intensities extending along $\langle110\rangle$ directions in reciprocal space (see Fig.~\ref{fig:1}),  which is sometimes denoted as the "butter-fly diffuse" [due to its shape in the (HK0) scattering plane] or "T2 diffuse" (due to its relation to T2 phonon modes in the system)~\cite{Zhijun}. In order to determine how the SRPO affect ferroelectric properties of these materials, it would be of great interest to study the response of the diffuse scattering to external electric field along different directions. In this paper we discuss our results of neutron diffuse scattering measurements on a single crystal of PMN-$28\%$PT  under an external field along [110] direction.

The single crystal of PMN-28\%PT is grown  by a modified Bridgman method at Penn State University.  It is located on the left side but very close to the MPB. In zero-field, the system undergoes a cubic-tetragonal-rhombohedral/monoclinic phase transition upon
cooling, with T$_{C1}\sim 440$~K and  T$_{C2}\sim 390$~K. The neutron diffuse scattering measurements have been carried out on SPINS cold triple-axis-spectrometer at the NCNR, with fixed E$_F$=5.0~meV, and collimations of Guide-80-80-open. Be filters are used both before and after the sample to reduce higher order
neutrons. Measurements have been performed in both the (HK0) and (H0L) planes while the external electric field is applied along the [110] direction.

\begin{figure}
    \includegraphics[width=0.9\linewidth]{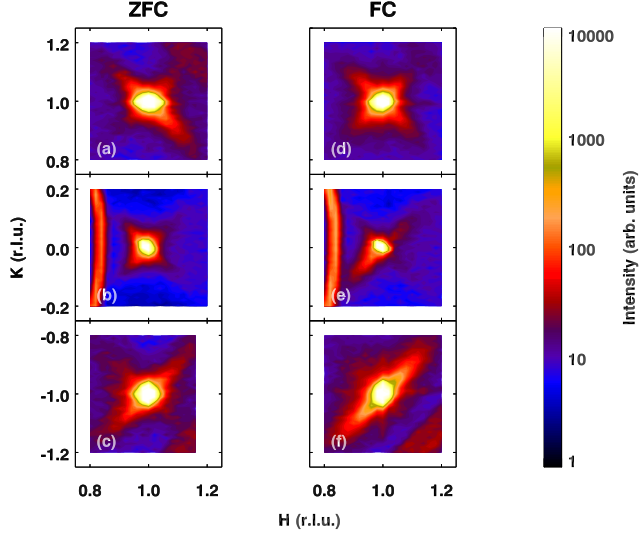}
    \caption{Mesh  intensity  maps measured at 300~K in the (HK0) plane near (110) (top row), (100) (center row) and (1$\bar{1}$0) (bottom row) Bragg peaks. The left column shows intensities measured under ZFC, while the right column shows measurements under FC of E=1~kV/cm along [110]. }\label{fig:2}
\end{figure}

The diffuse scattering intensity distributions under ZFC measured at 300~K are shown in the left column of Fig.~\ref{fig:2}. They are consistent with the known behavior of the T2-diffuse in PMN/PZN type relaxors where the intensity has a butterfly shape near (100) and are elongated along the transverse directions near (110) and (1$\bar{1}$0) Bragg peaks. The diffuse scattering intensity from the SRPO is a result of displacement type (short-range) order and  follows the $|{\bf Q}\cdot{\bf\epsilon}|^2$ factor where ${\bf Q}$ is the wave-vector transfer and ${\bf\epsilon}$ is the polarization vector (atomic shift). If one decomposes the T2-diffuse in the (HK0) plane into two "wings" (red and blue, shown in Fig.~\ref{fig:1}), it would be reasonable to associate the red "wing" with [110] type local polarizations, based on its being intense around ${\bf Q}=(110)$ and weak/absent around ${\bf Q}=(1\bar{1}0)$. Likewise, the blue "wing" intensity is naturally
associated with [1$\bar{1}$0] type polarizations~\cite{Xu_3D}. Under ZFC, in average these wings are equally intense near the (100) Bragg peak.

\begin{figure}
    \includegraphics[width=0.9\linewidth]{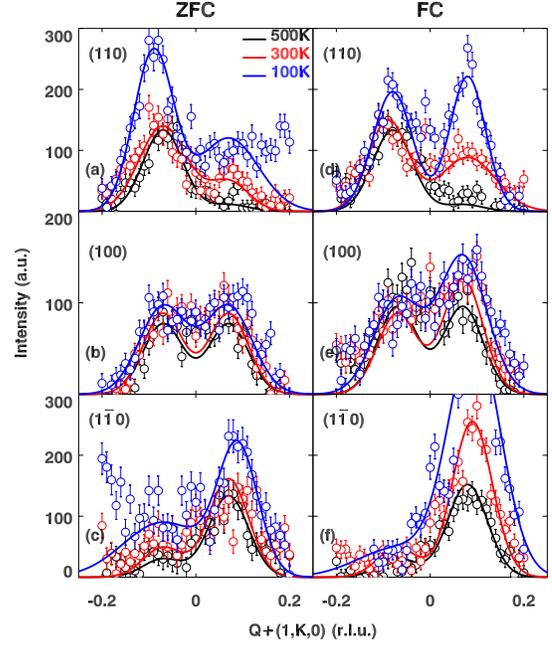}
    \caption{Linear intensity profiles measured along [1.1,K,0] at 100~K (blue), 300~K (red) and 500~K (black). The left column shows intensities measured under ZFC, while the right column shows measurements under FC of E=1~kV/cm along [110]. }\label{fig:3}
\end{figure}

A linear intensity profile taken along [1.1,K,0] (the dashed line in Fig.~\ref{fig:1}) can be used to monitor how these two "wings" change without having to complete the entire 2D intensity mesh. In the left column of Fig.~\ref{fig:3},
we show the temperature dependence of the diffuse scattering intensity along [1.1,K,0] upon ZFC. Because this dashed line
is taken on the right side (positive H side) of all three [(110),(100) and (1$\bar{1}$0)] Bragg peaks, the red
wing intensity is always going to show up on the left side (negative K side) of the blue wing intensity. From the data one can clearly see that the diffuse scattering intensity grows upon cooling - the growth is more pronounced near (110) and (1$\bar{1}$0) than (100) suggesting a change of the diffuse scattering structure
factors across different Bragg peaks upon cooling. As expected, the blue and red wing intensities develop
equally near (100). Near (110), the red wing (the left peak) is  much stronger,  and near (1$\bar{1}$0) the blue wing (the right peak) is dominating instead.

When an external field of E=$1~$kV/cm along [110] is applied at 500~K, mesh scans performed at 300~K
[Fig.~\ref{fig:2}(d), (e) and (f)]  indicate that intensities from the blue wing
($\epsilon_{blue} // [1\bar{1}0]$) are
enhanced by the field and those from the red wing ($\epsilon_{red} // [110]$)
are reduced. This is also
apparent when we investigate the linear intensity profiles along [1.1,K,0]  (Fig.~\ref{fig:3}), where the peak on the right side
is significantly enhanced with FC. It is worth noting that the enhanced blue wing intensity is present
even near the (110) Bragg peak, where ${\bf Q}$ is perpendicular to $\epsilon_{blue}$. There could be
two possible explanations, (i) the polarization of the SRPO that contribute to the blue wing intensity could
have been slightly affected by the field and therefore not entirely perpendicular to ${\bf Q}$ anymore;
or (ii) these SRPO are dominated by $\langle110\rangle$ type polarizations, but could still have a small portion of polarization components along other directions, and therefore the blue wing intensity is never completely extinct near (110).
The latter seems more plausible since that even without an external field, small traces of the
blue wing intensities can still be observed near (110) [see Fig.~\ref{fig:1}(a) and (c), also small traces of red wing intensities near (1$\bar{1}$0)].

\begin{figure}
    \includegraphics[width=0.9\linewidth]{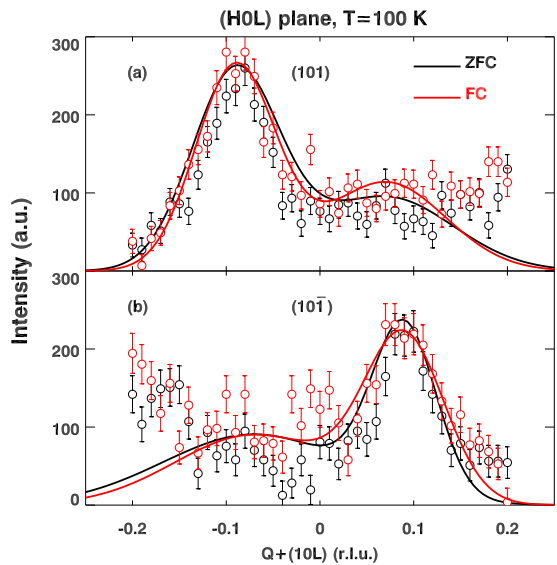}
    \caption{Linear intensity profiles measured along [1.1,0,L] at 100~K near (101) and (10$\bar{1}$). ZFC measurements are shown in black, while FC (E=1kV/cm along [110]) measurements are shown in red. }\label{fig:4}
\end{figure}

While in the (HK0) plane we observe this [110]-field induced
redistribution of diffuse scattering intensity from SRPO
with $\epsilon_{red} // [110]$ to $\epsilon_{blue} // [1\bar{1}0]$ in the low temperature phase, it is
important to perform similar measurements on diffuse scattering in (H0L) and/or (0KL) planes. Our findings
are that the diffuse scattering intensities in the low temperature R-phase are not affected by the [110] field in these
planes. An example of intensity profiles along [1.1,0,L] (in (H0L) plane) are shown in Fig.~\ref{fig:4}.
The field has no effect for diffuse scattering intensities near either the (101) or (10$\bar{1}$) Bragg peaks.
Suggesting that the [110] field does not affect diffuse scattering intensities from SRPO with $\epsilon$
along [101],[10$\bar{1}$] [measured in (H0L) plane], [011], and [01$\bar{1}$] [measured in (0KL) plane].

In Table~\ref{tab:1} we summarize the observed electric field effect on various T2-diffuse components associated
with different $\langle110\rangle$ polarizations. We find that the results are not sensitive to the strength
of the field
(moderate E-fields ranging from 0.5~kV/cm to 4~kV/cm all have similar effect). We also include results of [001] and [111] field reported in previous work~\cite{Xu_3D,Xu_coexist,Wen_PMN} for comparison. We notice that in general, one sees a trend that when the polarization of the SRO is perpendicular to the field, the associated diffuse scattering component is likely to be enhanced (e.g. the case of $\epsilon\parallel[1\bar{1}0]$ and E$\parallel[110]$, or $\epsilon\parallel [1\bar{1}0]$ and $E\parallel [111]$, etc.). This is however, not always the case, for example, for E$\parallel$[001], even when $\epsilon\parallel[110]$ which is perpendicular to E, no enhancement occurs.

\begin{table}
    \caption{Electric field induced intensity change of diffuse scattering associated with short-range polarizations with $\epsilon$ along different $\langle110\rangle$ directions. "NC" denotes "no change".}
    \begin{ruledtabular}
        \begin{tabular}{c|c|c|c|c|c|c}
            $\epsilon$ &  [110] & [1$\bar{1}$0] & [101] & [10$\bar{1}$] &[011] & [01$\bar{1}$]\\
            \hline
            E$\parallel$[110] & - & + & NC & NC & NC & NC\\
            \hline
            E$\parallel$[111] & - & + & - & + & - & +\\
            \hline
            E$\parallel$[001] & NC & NC & NC & NC & NC & NC\\

        \end{tabular}
    \end{ruledtabular}
    \label{tab:1}
\end{table}

These types of field effects on T2-diffuse scattering are never observed for temperatures greater than T$_C$. In addition
to the FC measurements, they can also be induced directly at low temperature by applying a field to the sample without having to go through a field cooling process through T$_C$. Moreover, these effects persist at low temperature even after the field is removed. This type of history dependence suggests a connection to the formation of ferroelectric domains. Being on the left side of the MPB, the low temperature ground state of PMN-28\%PT (and other PZN-$x$\%PT and PMN-$x$\%PT solid solutions with low PT concentrations) is in average rhombohedral, or, upon FC, monoclinic, that are slightly modified from the rhombohedral state. Therefore the polarizations of the ferroelectric domains are $\langle111\rangle$ (R-phase) or slightly rotated from $\langle111\rangle$ (M-phases).

The observed change of T2 diffuse scattering intensity distribution can only be explained if, within each of the
$\langle111\rangle$ ferroeletric domains, local regions with different $\langle110\rangle$ SRPO are not equally populated.
Consider the case of E-field along [111], the configuration in Table~\ref{tab:2} would be a natural solution. Here we propose that in ferroelectric domains with the four different $\langle111\rangle$ type polarizations (we ignore the positive/negative polarity), only local regions with SRPO perpendicular to the surrounding polarization can develop. If an external field along [111] is applied which can increase the volume of $P_{111}$,
consequently, diffuse scattering intensities associated with [1$\bar{1}$0],[10$\bar{1}$], and [01$\bar{1}$]
local polarizations are enhanced~\cite{Xu_coexist} (see Table~\ref{tab:1} too). In the case of an external field along [001]
direction, the field effect on the four domains (P$_{111}$, P$_{11\bar{1}}$, P$_{1\bar{1}1}$, and P$_{\bar{1}11}$) are the same and none of these is more favored than the other. Therefore one will not observe
any clear change of the T2-diffuse scattering intensity distribution.

In the current case, when a [110] field is applied, the situation is a bit more complicated. One would expect the P$_{1\bar{1}1}$ and P$_{\bar{1}11}$ domains to diminish since they have polarizations perpendicular to the
field and are not favored in energy during the domain formation process. The other two domains, P$_{111}$ and P$_{11\bar{1}}$ would have increased volumes (compared to the zero-field condition). The polarizations of these domains can of course, be rotated away slightly from the $[111]$ and $[11\bar{1}]$ directions in the corresponding monoclinic planes by the field, which  nevertheless, does not affect our discussion for the SRPO.
As a result, based on our proposed SRPO distribution in different ferroelectric domains in Table~\ref{tab:2},
volume of SRPO  along $[1\bar{1}0]$ will increase (present in both P$_{111}$ and P$_{11\bar{1}}$), and that of SRPO along $[110]$ will decrease (not present in either P$_{111}$ or P$_{11\bar{1}}$). The volume of
SRPO along $[101],[10\bar{1}],[011]$, and $[01\bar{1}]$ will not change (the volume increase from P$_{111}$ and P$_{11\bar{1}}$ and the volume decrease from  P$_{1\bar{1}1}$ and P$_{\bar{1}11}$ cancel each other out for these SRPO). This naturally explains the partial redistribution of diffuse scattering intensities observed in this study (Table~\ref{tab:1}).

\begin{table}
    \caption{The distribution of short-range polarizations with $\epsilon$ along  $\langle110\rangle$ directions in ferroelectric domains with different $\langle111\rangle$ polarizations."Y" and "N" denote whether such a SRPO can develop in the ferroelectric domain.}
    \begin{ruledtabular}
        \begin{tabular}{c|c|c|c|c|c|c}
            $\epsilon$ &  [110] & [1$\bar{1}$0] & [101] & [10$\bar{1}$] &[011] & [01$\bar{1}$]\\
            \hline
            P$_{111}$ & N & Y & N & Y & N & Y\\
            \hline
            P$_{11\bar{1}}$ & N & Y & Y & N & Y & N\\
            \hline
            P$_{1\bar{1}1}$ & Y & N & N & Y & Y & N\\
            \hline
            P$_{\bar{1}11}$ & Y & N & Y & N & N & Y\\

        \end{tabular}
    \end{ruledtabular}
    \label{tab:2}
\end{table}

The current results, when interpreted using this proposed picture, suggests that a moderate electric field mainly change the volumes of various ferroelectric domains, rather than affecting the SRPO directly. This reiterates results from other work showing that the SRPO are strongly dependent on the chemical short-range order~\cite{Burton2,Prosandeev2018}, where the latter is obviously not sensitive to a moderate electric field.
In addition, our field-measurements provides a way to understand how the SRPO and long-range ferroelectric order coexist, by tuning
the relative volumes of domains with different long-range polarizations.  The configuration  where the SRPO
develop with polarizations perpendicular to that of their surrounding ferroelectric domains may
at first appear aberrant since it is clearly not an energy-favorable state. We do not have a convincing explanation for why this configuration could occur. However, one may look at the problem from another
perspective. The SRPO can be observed in these relaxor compounds, because they differ from the surrounding
environment. If local SRPO develop with polarizations similar to the surrounding lattice matrix (in the low
temperature ferroelectric phase), they may eventually blend into the polar lattice and become hard to distinguish. On the other hand, local polarizations perpendicular to the surrounding lattice will always
remain distinguishable and stand out in any bulk measurements that are sensitive to local polarizations thatr
deviate from the lattice matrix.

Overall, we have shown that in PMN-28\%PT, a partial redistribution of T2-diffuse scattering intensity can be induced by a moderate electric field applied along [110] direction in the low temperature (R) phase.
Our results, together with previous results
on the full-redistribution of T2-diffuse scattering intensities under [111] field, and the lack of change of the
T2-diffuse scattering intensities under [001] field, can all be well accounted for if the SRPO in these relaxor
compounds are distributed in a configuration where they develop mainly with $\langle110\rangle$ polarizations
normal to the $\langle111\rangle$-type polarizations in the surrounding R-type ferroelectric domains. More work
on the SRPO in other phases, and theoretical considerations on why this configuration can be established in the
first place, are clearly needed.

\begin{acknowledgments} FL acknowledges the National Natural Science Foundation of China (Grants No. 51572214) and the Key Science and Technology project of Shanxi Province (2018KJXX-081). SZ acknowledges the support of ONRG under Grant No. N62909-18-12168. CS acknowledges the Carnegie Trust for the Universities of Scotland and the Royal Society.
\end{acknowledgments}


%

\end{document}